\documentclass[aps,prl,twocolumn,10pt,shortbibliography]{revtex4-1}
\usepackage{amsmath}
\usepackage{amsfonts}
\usepackage{amssymb}{\tiny }
\usepackage{mathtools}
\usepackage{times}
\usepackage{bm}
\usepackage{cases}
\usepackage[version=4]{mhchem}
\usepackage{graphicx}
\usepackage[caption=false]{subfig}
\usepackage[usenames]{color}
\usepackage[bookmarks=true,colorlinks,linkcolor=blue, urlcolor=blue,citecolor=blue]{hyperref}

\definecolor{darkred}{rgb}{0.847059, 0.141176, 0.164706}
\definecolor{darkgreen}{rgb}{0,0.4,0}
\definecolor{darkblue}{rgb}{0.254902, 0.411765, 0.882353}

\allowdisplaybreaks%
\newcommand{\bs}{\boldsymbol}
\newcommand{\mc}{\mathcal}
\newcommand{\mb}{\mathbb}
\newcommand{\dg}{\dagger}
\newcommand{\pg}{{\phantom{\dagger}}}

\begin{document}
\title{Topological phase transition and nontrivial thermal Hall signatures in \\ honeycomb lattice magnets}
\author{Yonghao Gao$^{1}$}
\author{Xu-Ping Yao$^{2}$}
\author{Gang Chen$^{2,1}$}
\email{gangchen.physics@gmail.com}
\affiliation{$^{1}$State Key Laboratory of Surface Physics and Department of Physics, Fudan University, Shanghai 200433, China}
\affiliation{$^{2}$Department of Physics and Center of Theoretical and Computational Physics, The University of Hong Kong, Pokfulam Road, Hong Kong, China}

\date{\today}

\begin{abstract}
We investigate spinon band topology and engineering from the interplay between long-ranged magnetic order and fractionalized spinons, as well as Zeeman coupling under external magnetic fields, in honeycomb lattice magnets. The synergism of N\'eel order and magnetic fields could reconstruct the spinon bands and drive a topological phase transition from the coexisting phase of long-ranged order and chiral spin liquid with semion topological order to the conventional magnetic order. Our prediction can be immediately tested through thermal Hall transport measurements among the honeycomb lattice magnets that are tuned to be proximate to the quantum critical point. Our theory should also shed light on the critical behavior of honeycomb Kitaev materials with emergent Majorana fermion bands. We suggest a possible relevance to 
the spin-1/2 honeycomb spin liquid candidate material In$_3$Cu$_2$VO$_9$. 
\end{abstract}
\maketitle

\emph{Introduction.}---Since the concept of resonated valence bond state was introduced by P. W. Anderson~\cite{Anderson1973}, great progress has been made to understand the quantum spin liquid (QSL), an exotic quantum state of matter characterized by fractionalized spin excitations and emergent gauge structures~\cite{Balents2010,RevModPhys.89.025003,Savary2016}. The description of the QSLs goes beyond the traditional Landau’s paradigm that defines phases from their local order parameters and symmetry-breaking patterns.  Historically, the original proposal of a QSL was on the geometrical frustrated triangular-lattice antiferromagnet, thus the search for QSL states in quantum magnets has mainly focused on the frustrated triangular, kagom\'e, pyrochlore lattice materials~\cite{Balents2010,RevModPhys.89.025003,Savary2016}. However, the geometrical frustration is not necessary, the essential ingredient to realize QSLs is the interplay between competing interactions and quantum fluctuations. A prominent example is the Kitaev spin-$1/2$ model on a honeycomb lattice, where geometrical frustration is absent~\cite{KITAEV20062}. Instead, it is the presence of bond-dependent Kitaev interactions that induces strong quantum fluctuations and frustrates spin orders. The Kitaev honeycomb model is exactly solvable and its ground state can be a gapped or gapless $\mb{Z}_2$ QSL depending on the relative strength of the Kitaev interactions along three different bonds~\cite{KITAEV20062,Wen2019}.  Jackeli and Khaliullin further laid out the essential ingredients for the realization of Kitaev model in Mott insulating iridates with spin-orbit-entangled local moments~\cite{PhysRevLett.102.017205}, which ignited the experimental synthesis of Kitaev materials and exploration of Kitaev QSL~\cite{Rau2016,PhysRevB.93.214431}.

\begin{figure}[b]
	\centering
	\includegraphics[width=6.5cm]{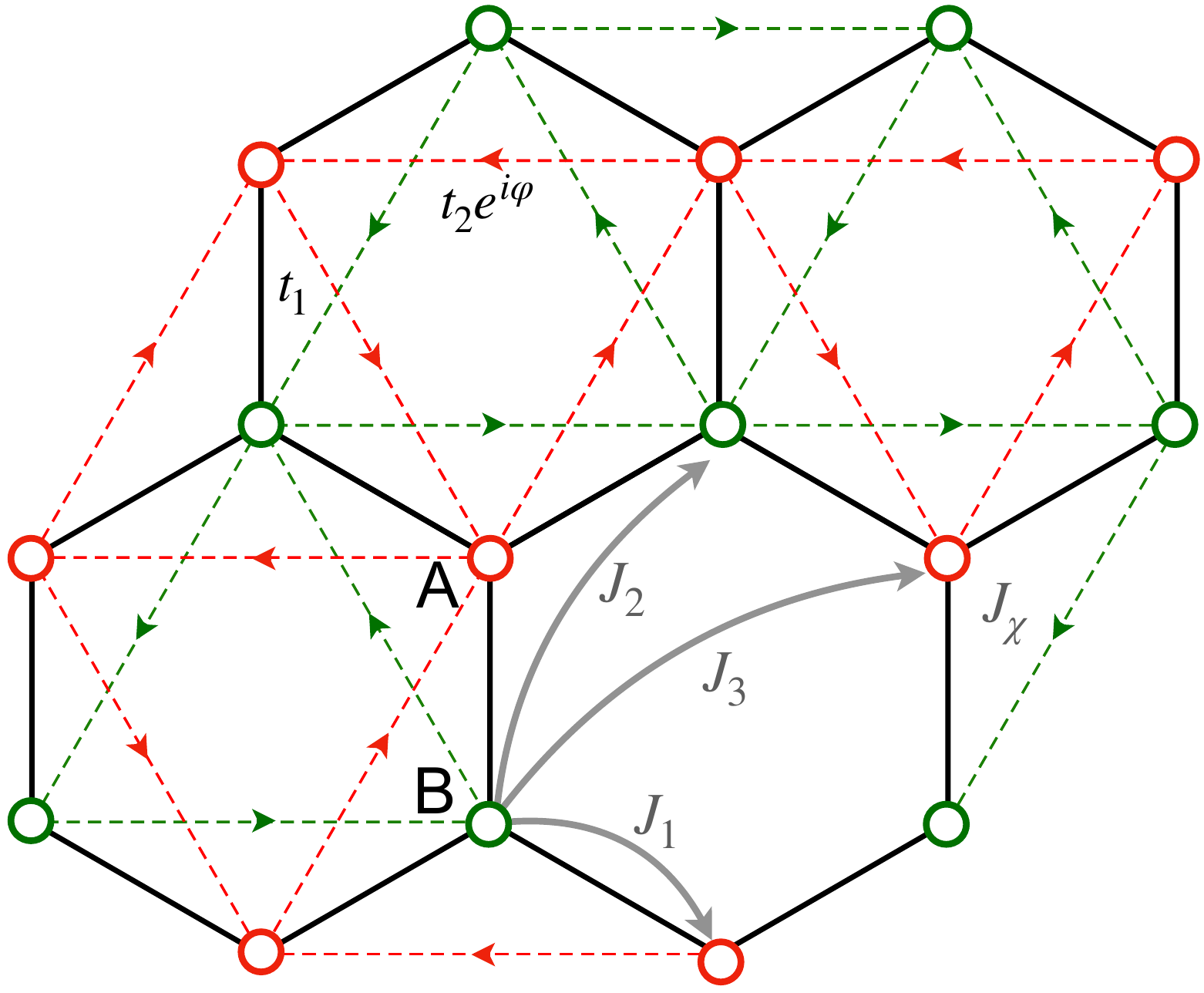}
	\caption{Schematic illustration of the hopping matrix up to second neighbors on a honeycomb lattice, 
	where for the nearest neighbor hopping ${t_{1,ij}=t_{1,ji}=t_1}$, and  
	for the second neighbor hopping ${t_{2,ij}=t_2e^{i\varphi}}$ 
	when the spinon hops along the (dashed) arrows or ${t_{2,ij}=t_2e^{-i\varphi}}$ 
	when the spinon hops oppositely the arrows. 
	The (light) gray curve arrows represent Heisenberg exchanges up to third neighbor, 
	while $J_{\chi}$ refers to the scalar spin chirality term related to three neighbor sites.}
	\label{fig1}
\end{figure}

Besides the Kitaev honeycomb model and the search for Kitaev materials, the antiferromagnetic $J_1$-$J_2$ spin-1/2 Heisenberg model on the honeycomb lattice has also attracted enormous attention since the second neighbor interaction could introduce a strong frustration into the system. It is generally believed that the ground state of the nearest-neighbor Heisenberg model on the honeycomb lattice is a conventional antiferromagnetic N\'eel order, while turning on the second-neighbor interaction would melt this long-range order and drive the system into a quantum disordered phase. In fact, a variety of numerical studies~\cite{PhysRevLett.107.087204,PhysRevB.88.165138,PhysRevLett.110.127205,PhysRevB.85.060402,Bishop2012,Liu2020} have suggested that the QSL phase could emerge from the spin-$1/2$ antiferromagnetic $J_1$-$J_2$ Heisenberg model on the honeycomb lattice for intermediate $J_2/J_1$, while the specific parameter range of it has been greatly debated and the detailed properties of the candidate QSLs have not yet reached a consensus. Remarkably, a very recent paper~\cite{Liu2020} found two topologically different phases in the intermediate disordered regime, one of which is the $\pi/2$-flux chiral spin liquid (CSL) with the semion topological order. In their case, the second neighbor exchange $J_2$ in the CSL already behaves the similar properties as the flux term in the Haldane model~\cite{PhysRevLett.61.2015}, and a large $J_2$ term promotes spinons to acquire a topological phase similar to the spin-orbital coupling in the Kane-Mele model~\cite{PhysRevLett.95.226801}. Beyond the pure $J_1$-$J_2$ Heisenberg model, Ref.~\cite{PhysRevLett.116.137202} further considered the third-neighbor exchange $J_3$ and the scalar spin chirality term $J_{\chi}$, and singled out a parameter window of the CSL proximate to the conventional N\'eel order. They formulated a gauge theory to study the transition from the CSL to another proximate confining tetrahedral state. 

\begin{figure}[t]
	\centering
	\includegraphics[width=8cm]{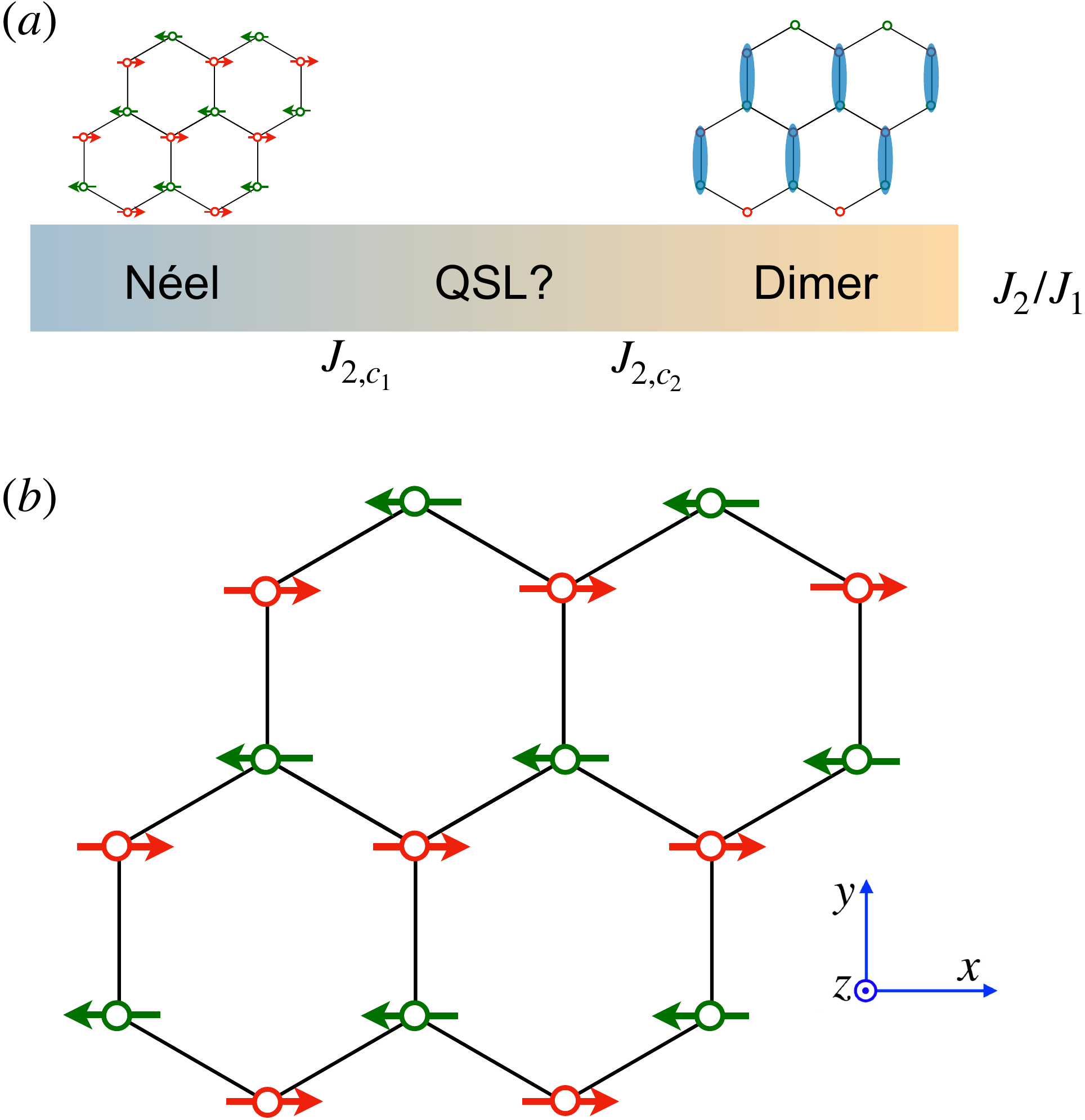}
	\caption{(a) General phase diagram from the numerical studies for a pure $J_1$-$J_2$ Heisenberg model on the honeycomb lattice.  For the small $J_2$ region, the ground state is generally believed to be a long-ranged N\'eel order, while $J_2$ becomes comparable to $J_1$, a dimer state or a stripe order could be stabilized, and the intermediate regime is proposed as a QSL, both gapped and gapless. (b) The N\'eel state. Here we choose the order along $x$-direction to minimize the  energy under a $z$-direction external magnetic field.}
	\label{fig2}
\end{figure}

In this work, instead of directly solving a specific spin model numerically on the honeycomb lattice and then determining the detailed properties of the intermediate quantum-disordered regime, we assume that the intermediate regime harbors a QSL phase and investigate the phase transition from a coexisting phase of QSL and N\'eel order to the conventional magnetic order under the external fields. Given the suggestion of a CSL~\cite{Liu2020}, we identify a topological phase transition with increasing magnetic fields. Especially, we find a quantized thermal Hall effect in the coexisting phase and a non-trivially enhanced thermal Hall conductivity in the confining ordered phase near the quantum critical point, similar to the discussion in the context of unusual thermal Hall effect for pseudogap phase of copper-based superconductors~\cite{Samajdar2019,Grissonnanche2019}. The situation that we considered here would apply to the relevant quantum materials with multiple competing phases, where the interplay among conventional ordered states, fractionalized elementary excitations in QSLs and Zeeman coupling together drive the topological phase transition and result in nontrivial thermal Hall signatures.

\emph{Spin model and parton construction.}---Although we do not attempt to solve any specific spin models, it would be very instructive to start from a general spin model on the honeycomb lattice for further investigations, from which we can clearly see where the degrees of freedom we considered could emerge. 
For concreteness, we begin with the following spin Hamiltonian on the honeycomb lattice,
\begin{equation}
\label{Eq: model}
H = \sum_{i<j} J_{ij}\bs{S}_i\cdot\bs{S}_j + J_{\chi}\sum_{i,j,k\in\triangle}\bs{S}_i\cdot\bs{S}_j\times\bs{S}_k,,
\end{equation}
where $ \bs{S}_i$ is the spin-$1/2$ operator at the site $i$, ${J_{ij} > 0}$ is the antiferromagnetic Heisenberg exchange, that can be extend to second neighbor, third neighbor and so on, as shown in Fig.~\ref{fig1}.  Although there is no geometrical frustration on honeycomb lattice, by switching on an antiferromagnetic 
$J_2$ term or further neighbor exchange would indeed bring competing interactions. An extremely important question is when the conventional N\'eel order is destroyed by the competing interactions and quantum fluctuations, what kind of states
 emerge from the melted phase. This question has long been pursued by a variety of numerical studies~\cite{PhysRevLett.107.087204,PhysRevB.88.165138,PhysRevLett.110.127205,PhysRevB.85.060402,Bishop2012,Liu2020}, but still without consensus on the exact properties of the intermediate phase. In Fig.~\ref{fig2} (a), we plot a general phase diagram of the $J_1$-$J_2$ Heisenberg model on honeycomb lattice. For the small $J_2$ region, just as the nearest-neighbor Heisenberg model on the honeycomb lattice, the ground state should be a long-ranged N\'eel order, while $J_2$ becomes comparable to $J_1$, a dimer state or a stripe order could be stabilized, and the intermediate regime is proposed as a QSL, both gapped and gapless. Moreover, we have also introduced a scalar spin chirality term $J_{\chi}$ in Eq.~\eqref{Eq: model}, that is helpful to realize a CSL. Although the recent numerical study~\cite{Liu2020} has shown that a pure $J_1$-$J_2$ Heisenberg model on the honeycomb lattice is already able to realize a CSL, here we add it for further convenience and general discussion. The scalar spin chirality term $J_{\chi}$ breaks the time reversal symmetry $\mc{T}$ and parity $\mc{P}$, but preserves their combination $\mc{P}\mc{T}$. Physically, in the weak Mott insulators with strong charge fluctuations, the ring exchange process would lead to the coupling~\cite{PhysRevB.51.1922,PhysRevB.73.155115,PhysRevLett.104.066403} between the scalar chirality and external magnetic fields through Zeeman coupling as
\begin{equation}
-\frac{24t^3}{U^2}\sum_{i,j,k\in\triangle}\sin{\Phi}\bs{S}_i\cdot\bs{S}_j\times\bs{S}_k,
\end{equation}
that is derived from the higher-order perturbation theory of the Hubbard model. 
Here $\Phi$ is the magnetic flux through the triangular plaquette $\triangle$ in an anticlockwise way. 
For the strong Mott insulator with large charge gap, the interplay between the symmetry allowed second neighbor Dzyaloshinskii-Moriya (DM) interaction and Zeeman Coupling can induce~\cite{PhysRevB.87.064423,Gao2020} a scalar chirality proportional to the magnetic field $B$ and DM strength $D_z$ as
\begin{equation}
\bs{S}_i\cdot\bs{S}_j\times\bs{S}_k \propto D_zB.
\end{equation}
Both cases need a finite magnetic field to induce the scalar chirality, while the latter case depends on the orientation of the DM vector. Since we are considering the field-driven phenomena, the $J_{\chi}$ term introduced in Eq.~\eqref{Eq: model} is well justified. Moreover, starting from the Haldane-Hubbard model can naturally lead to the $J_{\chi}$ term without further applied fields~\cite{PhysRevB.37.9753,PhysRevLett.116.137202}.

\begin{figure}[t]
	\centering
	\includegraphics[width=8.5cm]{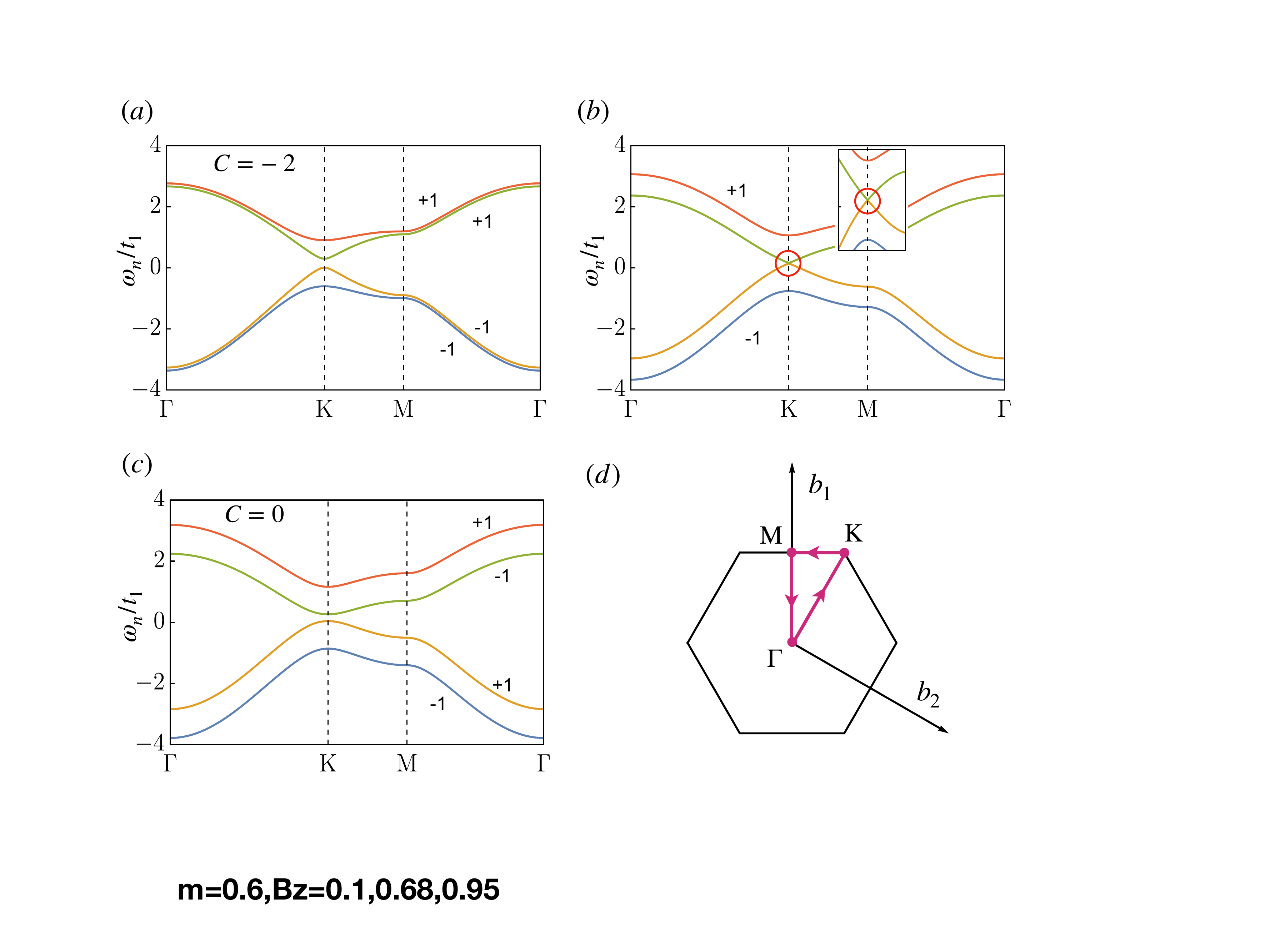}
	\caption{Representative spinon bands along the high symmetry momentum direction the Brillouin zone. The numbers $\pm 1$ near the bands stand for the corresponding Chern numbers, and the number $C$ represents the total Chern of the fully occupied spinon bands. In the calculation we have fixed ${\varphi=\pi/3}$, ${m/t_1=0.6}$ and ${t_2/t_1=0.1}$ while varying the magnetic fields for (a) ${B_z/t_1=0.1}$ (b) ${B_z/t_1=0.67}$ and (c) ${B_z/t_1=0.95}$. With the increasing of magnetic fields, the spinon bands experience a gap closing and reopening. (d) First Brillouin zone of honeycomb lattice and the high symmetry line marked by colored arrows, $b_1$ and $b_2$ are two basis vectors of the reciprocal lattice.}
	\label{fig3}
\end{figure}

\begin{figure*}[htp]
	\centering
	\includegraphics[width=15cm]{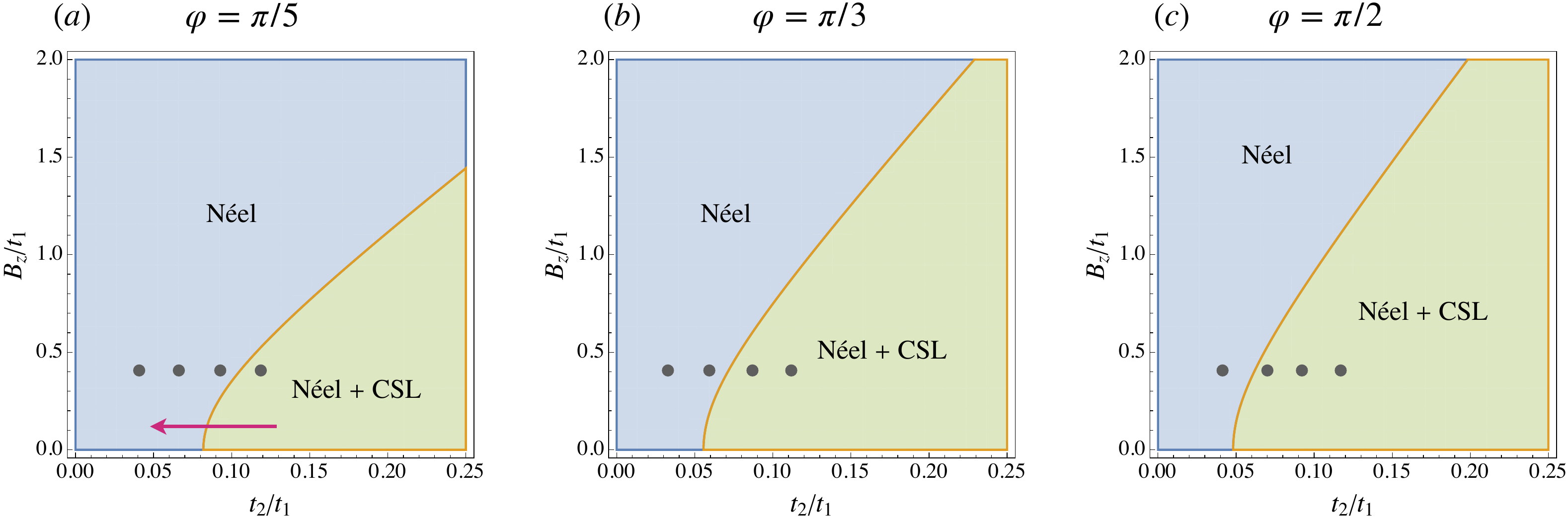}
	\caption{Mean-field phase diagram for three certain phases $\varphi$ with varying second neighbor hopping coefficient $t_2$ and magnetic field $B_z$, while $m/t_1$ is fixed as $1/2$ here. (a) ${\varphi=\pi/5}$, (b) ${\varphi=\pi/3}$ and  (c) ${\varphi=\pi/2}$. Specifically, the phase ${\varphi=\pi/2}$ corresponds to a pure imaginary second neighbor hopping coefficient. The colored arrow in (a) represents for a phase transition from the coexisting phase of magnetic order and CSL to the conventional antiferromagnetic N\'eel state, well compatible with the fact that the second neighbor exchange brings the competing interaction, and decrease of $t_2$ would recover the conventional magnetic order. The (dark) gray dots in the phase diagrams correspond the parameters we have chosen to calculate the thermal Hall conductivity later.}
	\label{fig4}
\end{figure*}

To describe the QSL with the fractionalized excitations, we here adopt the Abrikosov fermion construction for the physical spin operator, which is one of the convenient parton approaches to study the QSL physics. In the Abrikosov fermion representation, the effective spin operator $\bs{S}_{i}$ on site $i$ is given by ${\bs{S}_{i}^{} =\frac{1}{2}\sum_{\alpha \beta} f_{i\alpha}^\dagger \bs{\sigma}_{\alpha\beta}^\pg f_{i\beta}^\pg}$, with $f_{i\alpha}$ (${\alpha = \uparrow, \downarrow}$) being the fermionic
spinon operator and $\bs{\sigma}$ being a vector of three Pauli matrices.  
The Hilbert space is enlarged due to the introduction of spinons, thus the constraint
${\sum_\alpha f^\dg_{i\alpha} f^\pg_{i\alpha} = 1}$ on local fermion number is imposed to project out unphysical states. Substituting the fermion representation into the spin Hamiltonian Eq.~\eqref{Eq: model}, one would obtain an interacting fermion system, which is an exact representation of the original model  
with the local occupation constraint, but still remains unsolvable. To tackle the reformulated interacting fermionic system, a useful and convenient way is to perform a quadratic decoupling~\cite{PhysRevB.65.165113} and recast the spion Hamiltonian into a generic quadratic form,
\begin{equation}
\label{Eq: Ham}
\begin{split}
H_{\rm QSL}=&-\sum_{i < j,\alpha\beta}(t_{ij}^{\alpha\beta}f_{i,\alpha}^{\dagger}f_{j,\beta}+\Delta_{ij}^{\alpha\beta}f_{i,\alpha}^{\dagger}f_{j,\beta}^{\dagger}+h.c.)\\
&-\sum_{i,\alpha}\mu_i f_{i,\alpha}^{\dagger}f_{i,\alpha},
\end{split}
\end{equation}
where the parameter $t_{ij}^{\alpha\beta}$ corresponds to spinon hopping channel 
while $\Delta_{ij}^{\alpha\beta}$ corresponds to spinon pairing channel between sites $i,j$, and 
the local chemical potential $\mu_i$ is introduced as a Lagrange multiplier to enforce Hilbert space constraint. Generally, $t_{ij}^{\alpha\beta}$ and $\Delta_{ij}^{\alpha\beta}$ should involve strong phase and amplitude fluctuations, and only the state that could survive against gauge fluctuations can be a deconfined QSL~\cite{PhysRevB.65.165113}.

The fermionic spinon carries spin-1/2 but does not have conventional electrical charge, thus it only couples to the external magnetic field through a linear Zeeman coupling,
\begin{equation}
H_B=-\frac{B_z}{2}\sum_{i,\alpha \beta} f_{i,\alpha}^{\dagger}
\sigma^z_{\alpha \beta}f_{i,\beta},
\end{equation}
where we have taken the $z$-direction external field for concreteness, and the Bohr magneton $\mu_B$ and Land\'e $g$ factor have been absorbed in $B_z$. It is already quadratic and does not need further decoupling.

In the coexisting phase of the quantum disordered QSL and the long-ranged N\'eel order, a moderate Zeeman coupling would minimize the energy of the honeycomb lattice antiferromagnet by tuning the N\'eel order to be orthogonal to the external magnetic field. Without loss of any generality, we fix the N\'eel order 
along the $x$-direction throughout this work under the external magnetic field along $z$-direction. 
Now we can consider the coupling between the conventional ordered spins and the fractionalized 
elementary excitations in the QSL as 
 \begin{equation}
 H_{\rm coupling}=\frac{m}{2}\sum_{i,\alpha \beta} \nu_i f_{i,\alpha}^{\dagger}
 \sigma^x_{\alpha \beta}f_{i,\beta},
 \end{equation}
where $m$ is magnetic component along $x$-direction and the factor $\nu_i$ takes $+1/-1$ for two different sublattices A/B, due to the staggered N\'eel order as shown in Fig.~\ref{fig2} (b). 
This is essentially a conventional order-parameter mean-field decoupling and is quadratic. 
With $z$ direction magnetic field, the N\'eel order orients in the $xy$ plane and is chosen to be along $x$  
in Fig.~\ref{fig2} (b). 
We ignore the fluctuations of the N\'eel order throughout this work 
as the magnon contribution does not influence our main result.

\emph{Mean-field analysis and phase diagram.}---We specifically choose the QSL to be a CSL with a semion topological order. It has been numerically demonstrated that this state could be stabilized in the honeycomb magnets, both for the pure antiferromagnetic $J_1$-$J_2$ Heisenberg model~\cite{Liu2020} and the extended spin model involving a finite third neighbor exchange and scalar spin chirality term~\cite{PhysRevLett.116.137202}. Additionally, it has been shown that the CSL can emerge in the Kitaev-$\Gamma$ model on honeycomb lattice with certain fields~\cite{PhysRevLett.120.187201}. Historically, Kalmeyer and Laughlin first proposed the CSL on the triangular lattice~\cite{PhysRevLett.59.2095}, that is closely related to the celebrated Laughlin wavefunction of the fractional quantum Hall effect. X.-G. Wen later identified~\cite{PhysRevB.40.7387} Chern-Simons theory as a topological field theory description of this chiral state. Recently, it has also been shown numerically that the CSL can be the ground state of several extended Heisenberg models on the kagom\'e lattice and on the triangular lattice with a non-zero $J_{\chi}$ interaction. To capture the CSL on the honeycomb lattice at the mean-field level, we proceed by decoupling the spin Hamiltonian Eq.~\eqref{Eq: model} to the Abrikosov fermion form in Eq.~\eqref{Eq: Ham} and further suppress the gauge fluctuations. Without the spinon pairing, one can simply ignore $\Delta_{ij}$ terms and only preserve the hopping sector. Moreover, in the mean-field treatment the local fermion occupation constraint can be replaced by the relaxed one, i.e., ${\sum_\alpha \langle f^\dg_{i\alpha} f^\pg_{i\alpha} \rangle = 1}$, then one could obtain a general quadratic spinon Hamiltonian with an uniform chemical potential $\mu$ and suppressed gauge fluctuations, that is given as follows 
\begin{equation}
\label{Eq: Ha}
\begin{split}
H_{\rm MF}=&-\sum_{i < j,\alpha}(t_{1,ij}f_{i,\alpha}^{\dagger}f_{j,\alpha}+t_{2,ij}f_{i,\alpha}^{\dagger}f_{j,\alpha}+h.c.)\\
&-\mu\sum_{i,\alpha} f_{i,\alpha}^{\dagger}f_{i,\alpha}.
\end{split}
\end{equation}
The amplitudes $t_{1,ij}$ and $t_{2,ij}$ are constrained by the corresponding projective symmetry group 
since the spinons fulfill the projective symmetries of the honeycomb lattice~\cite{PhysRevB.65.165113}. 
We choose a simple case and the value of $t_{ij}$ we taken is schematically depicted in Fig.~\ref{fig1}, 
where for the nearest-neighbor hopping ${t_{1,ij}=t_{1,ji}=t_1}$, and  for the second-neighbor hopping 
${t_{2,ij}=t_2e^{i\varphi}}$ when the spinon hops along the arrows and ${t_{2,ij}=t_2e^{-i\varphi}}$ 
when the spinon hops oppositely the arrows. The corresponding phase $\varphi$ could arise either 
from the decoupling of $J_{\chi}$ or just as a CSL ansatz of the pure Heisenberg model, 
that will be treated as a tuning parameter. Moreover, the chemical potential $\mu$ 
included to impose the Hilbert space constraint on average results in half-filling spinon bands. 
Then the full spinon Hamiltonian is
\begin{equation}
\label{Eq: hamMF}
H_{\rm total}= H_{\rm MF}+H_{B}+H_{\rm coupling}.
\end{equation}

% The spinon band evolution with applying magnetic is depicted in
% Fig.~\ref{fig3}. One can see that with the increasing of magnetic fields, the spinon bands experience a gap closing and reopening.

% We establish the phase diagram by determining the phase boundary when gap is closing, the results are depicted in Fig.~\ref{fig4}.

\begin{figure*}[t]
\centering
\includegraphics[width=15.8cm]{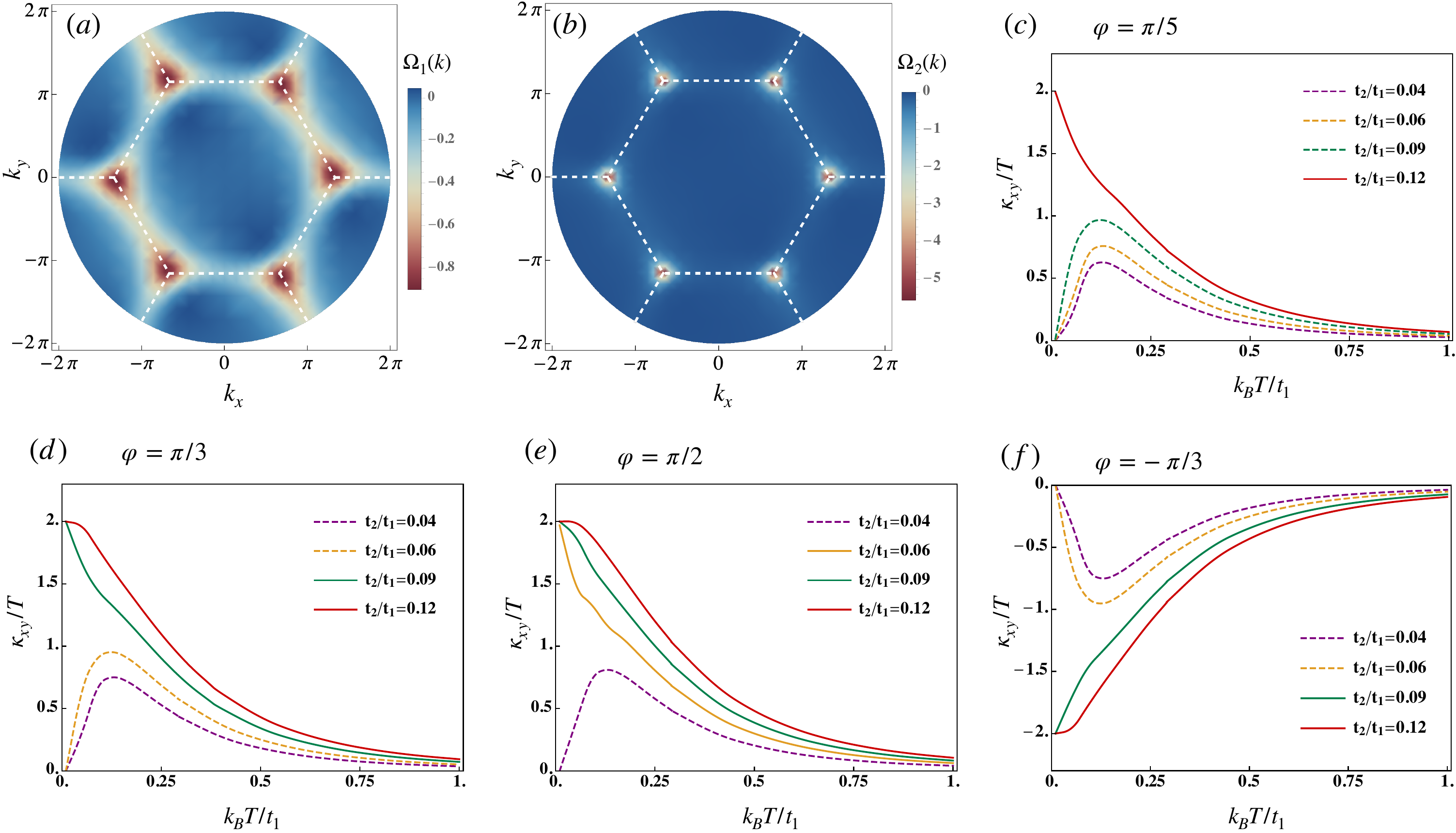}
\caption{Density plot of Berry curvatures calculated with ${t_2/t_1=0.1}$, ${B_z/t_1=0.4}$, ${m/t_1=1/2}$ 
and ${\phi=\pi/2}$ for (a) the lowest occupied spinon band and (b) the second occupied spinon band. 
The temperature dependence of thermal Hall conductivity with colored solid lines representing the 
thermal Hall conductivity in the coexisting phase with semion topological order and dashed lines 
standing for thermal Hall response in proximate confined ordered phase. 
The data are calculated with fixed ${m/t_1=1/2}$ and ${B_z/t_1=1/2}$, 
	while varying the temperature and $t_2$ for 
	(c) ${\varphi=\pi/5}$, (d) ${\varphi=\pi/3}$, (e) ${\varphi=\pi/2}$ and (f) ${\varphi=-\pi/3}$. 
	The unit of $\kappa_{xy}/T$ here is $\pi k_B^2/6\hbar$.}
	\label{fig5}
\end{figure*}

In the mean-field analysis, we depict the spinon band evolution in Fig.~\ref{fig3} with various magnetic fields. In the absence of external magnetic field, the influence of long range N\'{e}el order is transmitted into fractionalized spinon degree of freedom through $H_{\text{coupling}}$ term and splits both the occupied and unoccupied spinon bands around $K$ point of Brillouin zone, as shown in Fig.~\ref{fig3} (a) (here to obtain well-defined Chern numbers we have applied a very weak magnetic field). A sufficiently large $t_2$ can stabilize the CSL coexisting with the N\'{e}el order. At the mean-field level, this phase is characterized by the vanishing spinon Fermi surface and non-zero total Chern number of the occupied spinon bands. Then the Chern-Simons term enters the theory for U(1) gauge fluctuations and results in a topological quantum field theory, corresponding to a semion topological order. With the increasing of magnetic fields, the spinon bands experience a gap closing and reopening [see Figs.~\ref{fig3} (b) and (c)]. Although the spinon bands separately  have well-defined and non-vanishing Chern numbers, the net Chern number of the occupied bands turns out to be 0, corresponding to a compact U(1) gauge theory in 2D. The gapped spinons can be integrated out, resulting in a pure compact U(1) gauge field that is always confined in 2D due to the proliferation of instantons~\cite{POLYAKOV1977429}, and the system enters a trivial state. A topological quantum phase transition occurs here since the net Chern number jumps from $-2$ to 0, indicating a transition from the topologically ordered state to a phase with a trivial topology. Thus, the external magnetic field drives the system from a nontrivial coexisting phase into a conventional N\'{e}el state.

We establish the phase diagram with distinct $\varphi$ and fixed ${m=1/2}$ by tracing the changes of spinon band gap and corresponding Chern numbers. The results are depicted in Fig.~\ref{fig5}. While the magnetic field can drive a phase transition as we have discussed above, decreasing the second-neighbor hopping $t_2$ can also diminish the interaction competition and then recover the conventional N\'{e}el order. 
We note that the approach is not self-consistent because the coupling between the magnetic field and the ordered spins is not involved here. A finite external magnetic field along $z$-direction would induce a 
non-zero magnetization in the same direction. 
However, this modification can be treated as an effective 
in-plane magnetization we have used in our model. In the weak field regime, the induced 
out-plane magnetization can be considered small enough such that the coupling between it and the spinon excitations in QSL could be ignored safely. Therefore, low field intervals in 
the phase diagrams are fairly reliable and there is no impact on our main conclusion in this work.

\emph{Nontrivial thermal Hall signatures.}---Experimentally, inelastic neutron scattering (INS) measurement is better to directly detect the magnetic excitations in spin systems, which reveals the sharp magnon excitations and two-spinon continuum in the spectrum. In contrast, thermal transport is more sensitive to probe the character of low-energy itinerant excitations, especially the thermal Hall transport may get rid of the phonon interference. Compared with INS measurement, thermal Hall transport even has the ability to reflect the topological properties of spinon bands, while the former only encodes the dynamical information of magnetic excitations. Actually, the pioneering work~\cite{PhysRevLett.104.066403} about thermal Hall effect in magnets by Katsura \emph{et al.} has stimulated intensive related studies both experimentally and theoretically~\cite{PhysRevB.91.125413,PhysRevB.99.165126,Gao2020,PhysRevResearch.1.013014,PhysRevResearch.2.013066,Watanabe8653,PhysRevLett.121.097203,PhysRevLett.115.106603,PhysRevLett.120.217205}. The magnon contribution and the possible spinon contribution to the thermal Hall effect have been observed in a series experiments~\cite{Watanabe8653,PhysRevLett.121.097203,PhysRevLett.115.106603,PhysRevLett.120.217205}. In particular, half-integer quantized thermal Hall effect proposed for Majorana fermions has also been reported~\cite{Kasahara2018} in the honeycomb Kitaev materials $\alpha$-RuCl$_3$, which, if confirmed, would be a revolutionary discovery of the Kitaev QSL.

To utilize this powerful experimental probe to examine the topological quantum phase transition and its critical behavior, we next explicitly demonstrate the finite thermal Hall conductivity in the coexisting phase of long-range magnetic order and CSL, and in the proximate confined ordered phase. The thermal Hall conductivity formula for a general non-interacting fermionic system with chemical potential $\mu$ is given~\cite{PhysRevLett.107.236601} as
\begin{equation}
\label{thermcon}
\kappa_{xy}=-\frac{k_B^2}{T}\int d\epsilon(\epsilon-\mu)^2
\frac{\partial f(\epsilon,\mu,T)}{\partial \epsilon}\sigma_{xy}(\epsilon)\,.
\end{equation}	
Here ${f(\epsilon,\mu,T)=1/[e^{(\epsilon-\mu)/k_BT}+1]}$ is the usual Fermi-Dirac distribution function, 
and ${\sigma_{xy}(\epsilon)}$ is the zero-temperature Hall coefficient 
for a system with the chemical potential $\epsilon$. It is defined by ${\sigma_{xy}(\epsilon) = - \frac{1}{\hbar}\sum_{\bs{k},\xi_{n,\bs{k}}<\epsilon}\Omega_{n,\bs{k}}}$ with the Berry curvature  $\Omega_{n\bs{k}}$ 
for the fermion band indexed by $n$, and the sum runs over all the Berry curvatures below the Fermi energy. 
In the zero-temperature limit, Eq.~\eqref{thermcon} recovers~\cite{Gao2020} the Wiedemann-Franz law 
and gives
\begin{equation}
\frac{\kappa_{xy}}{T}=-\frac{\pi k_B^2}{6\hbar}\sum_{n\in{\rm filled}}C_n,
\end{equation}	
since here $\mu$ lies in the gap, and $C_n$ is the Chern number of the $n$-th spinon band defined by ${C_n=\frac{1}{2\pi}\int_{\rm BZ}\Omega_{n,\bs{K}}}$. The typical density plot of Berry curvatures for the two occupied spinon bands in the coexisting phase are plotted in Figs.~\ref{fig5} (a) and (b), one can see the Berry curvatures most locate around the corner $K$ point of the Brillouin zone, especially the Berry curvature of the second band exhibits sharp peaks at $K$ points.

In Figs.~\ref{fig5} (c)-(e), we numerically calculate the temperature dependence of thermal Hall conductivity with the parameters marked by dark gray dots in phase diagrams Figs.~\ref{fig4} (a)-(c). In these figures, the colored solid lines represent the thermal Hall conductivity in the coexisting phase with a semion topological order, which is quantized to 2 in the zero temperature limit and decrease monotonically with increasing temperature. Finally, the vanishing value in the higher temperature region is consistent with the fact that the total Chern number of the spinon bands is 0. On the other hand, the dashed lines represent the thermal Hall conductivity in the proximate confined phase, which is exactly 0 in the zero temperature limit, but it increases rapidly with temperature and then decreases gradually after reaching a maximum in the finite-temperature regime. We note that the thermal Hall conductivity in the coexisting phase is quantized as expected, but the non-quantized and finite thermal Hall conductivity of the proximate confined phase with the same order of magnitude in the finite temperature region is rather nontrivial, since the magnon picture from the ordered phase only gives rise to a much smaller thermal Hall conductivity. This implies that the ordered phase near the topological state can result in a nontrivial thermal Hall signature due to the proximity effect of topological quantum critical point. The sign influence of the phase $\varphi$ is depicted in Fig.~\ref{fig5} (f), where we plot the temperature dependence of thermal Hall conductivity when ${\varphi=-\pi/3}$ with other parameters same as in Fig.~\ref{fig5} (d). One can see the only change is that the thermal Hall response also acquires a minus sign, which can be traced back to the Chern number exchanges between the occupied and unoccupied bands.

\begin{figure}[t]
 	\centering
 	\includegraphics[width=7.2cm]{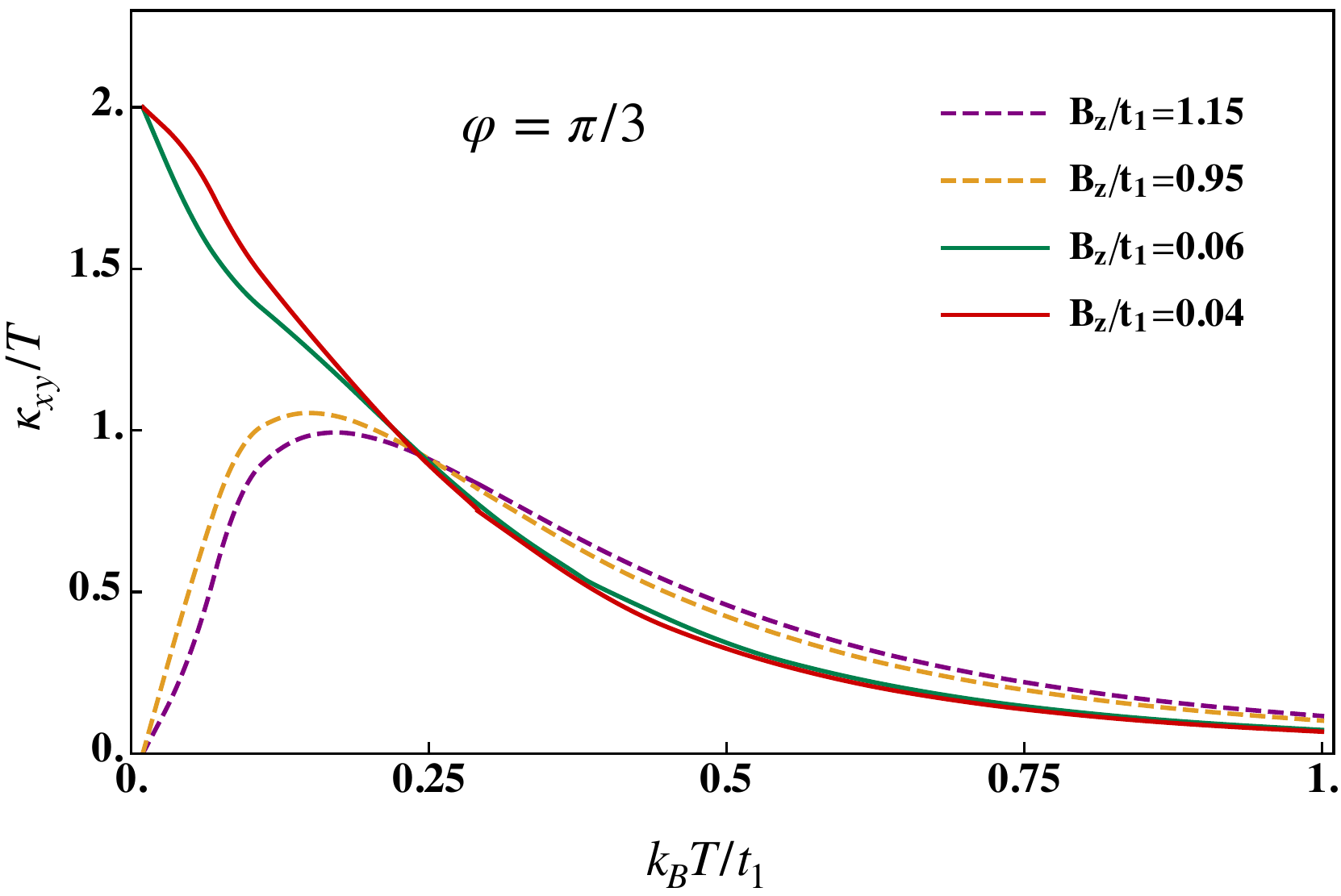}
 	\caption{The temperature dependence of thermal Hall conductivity calculated with fixed $m/t_1=1/2$ and second neighbor hopping amplitude $t_2/t_1=0,09$, while varying the temperature and magnetic field $B_z$ for $\varphi=\pi/3$. The unit of $\kappa_{xy}/T$ here is also $\pi k_B^2/6\hbar$.}
 	\label{fig6}
 \end{figure}

To further observe the field-driven transition, Fig.~\ref{fig6} displays the temperature dependence of thermal Hall response under four different magnetic fields $B_z$, where ${\varphi=\pi/3}$ and other parameters are fixed as explained in the caption. The main conclusion is very similar to that from Figs.~\ref{fig5} (c)-(e), while we  note that the thermal Hall conductivity curves cross in the finite-temperature region, which is slightly different from that in Figs.~\ref{fig5} (c)-(e) with well-separated curves, and this is nothing but a specific dependence on the band evolution under fields.

\emph{Discussion}---In conclusion, we have investigated the phase transition from a coexisting phase of QSL and N\'eel order to the conventional magnetic order under external fields. For the CSL, we identify a topological phase transition with increasing magnetic field, especially, we find a quantized thermal Hall effect in the coexisting phase and a nontrivial thermal Hall response in the confining ordered phase near the quantum critical point. The interplay between the conventional long-ranged magnetic order and Zeeman coupling is transmitted into the spinon bands and influence their topology. From the point of view of a pure band theory, the mathematical structure behind, in a certain sense, might be very similar to the celebrated Haldane model or its extension Kane-Mele model~\cite{PhysRevLett.61.2015,PhysRevLett.95.226801}, but the physical contents are fundamentally different. In the Haldane model or Kane-Mele model, they mainly focus on the single electron physics and the topology of corresponding electron wavefunction. While in our case, QSL is an emergent phenomenon from the strongly correlated electron system and its low-energy physics is effectively described by a compact gauge theory. In particular, when the spinon band is gapped and owns non-vanishing net Chern number, the Chern-Simons term enters the theory for gauge fluctuations and results in a topological quantum field theory. The corresponding quantum critical behavior could be very exotic and rather nontrivial. Similar physics has also been identified in Ref.~\cite{Samajdar2019} where they started from a $\pi$-flux QSL on square lattice and studied the proximity behavior of critical point to explain the experimental observation of giant thermal Hall conductivity in the pseudogap phase of cuprate superconductors~\cite{Grissonnanche2019}.

As for the specific material, In$_3$Cu$_2$VO$_9$~\cite{PhysRevB.85.085102} has been synthesized 
and the Cu ions form a honeycomb lattice with spin-1/2 local moments. The system probably develops 
a QSL ground state, though no strong evidence has been provided~\cite{PhysRevB.85.085102}.
Further first-principle calculation suggests frustrated spin interaction. In addition to further neutron study, It will be interesting to examine the magnetic field response and thermal transport in this system. Our theory 
may find an application in this compound. Furthermore, 
our result could apply to the Kitaev honeycomb lattice magnets with strong Kitaev interactions. Among the honeycomb Kitaev materials, so far, most of the them experience a phase transition to long-ranged magnetic order at low temperatures, such as the zig-zag order in $\alpha$-RuCl$_3$. Thus it would be very interesting to study the coexisting phase of magnetic order and Kitaev QSL under fields, which might tell us how the interplay of these degrees of freedom influence the topology of Majorana fermion bands and related critical behavior.  
 
Overall, we have considered here the honeycomb magnets with multiple competing phases, where the interplay between conventional ordered state and fractionalized spinon excitations in QSL, as well as a linear Zeeman coupling, together drives the topological phase transition and results in nontrivial thermal Hall signatures. It is rather appealing to investigate the coexisting phase of conventional magnetic ordered state 
and quantum disordered state, and the corresponding quantum critical behavior. 
Further works may involve the charge degrees of freedom, that might help us understand the relation between microscopic objects and macroscopic phenomena, for example, the high-temperature superconductivity.

\emph{Acknowledgments}---This work is supported by the Ministry of Science and Technology of China with Grant No. 2018YFGH000095, 2016YFA0301001, 2016YFA0300500, and by the Research Grants Council of Hong Kong with General Research Fund Grant No.17303819.

\bibliography{Ref.bib}

\end{document}